\documentclass[%
aip,
cp,  
 amsmath,amssymb,
 reprint,%
]{revtex4-1}

\usepackage[dvipdfmx]{graphicx}
\usepackage{dcolumn}
\usepackage{bm}
\usepackage{color}

\usepackage[utf8]{inputenc}
\usepackage[T1]{fontenc}
\usepackage{mathptmx}

\begin{document}

\title{Ferromagnetically ordered metal in the single-band Hubbard model}

\author{Akihisa Koga}
\affiliation{Department of Physics, Tokyo Institute of Technology, Meguro, Tokyo 152-8551, Japan}
\email[Corresponding author: A. Koga~]{koga@phys.titech.ac.jp}

\author{Yusuke Kamogawa} 
\affiliation{Department of Physics, Tokyo Institute of Technology, Meguro, Tokyo 152-8551, Japan}

\author{Joji Nasu}%
\affiliation{
   Department of Physics, Yokohama National University, 79-5 Tokiwadai, Hodogaya, Yokohama 240-8501, Japan
}

\date{\today} 

\begin{abstract}
  We study a ferromagnetic instability in a single-band Hubbard model
  on the hypercubic lattice away from half filling.
  Using dynamical mean-field theory with
  the continuous-time quantum Monte Carlo simulations based on the segment algorithm,
  we calculate the magnetic susceptibility
  in the weak and strong coupling regions systematically.
  We then find how ferromagnetic fluctuations are enhanced when
  the interaction strength and density of holes are varied.
  The efficiency of the double flip updates in the Monte Carlo simulations
  is also addressed.
\end{abstract}

\maketitle

Ferromagnetism in a metallic state has attracted much interest
since rare-earth-based permanent magnets has recently been synthesized such as
Nd-Fe-B.
In contrast to the multiorbital systems such as manganites,
the ferromagnetic instability in single-band models is less understood.
When the static mean-field approximation is applied to the model,
one meets the Stoner criterion, where
the Coulomb interaction yields a ferromagnetic instability
in the system with a large density of states (DOS) at the Fermi level.
The criterion is qualitatively correct when the interaction strength is small.
In fact, the ferromagnetically ordered states is realized in the single band systems
with flat bands~\cite{Mielke1,Mielke2,Mielke3,Tasaki,Kusakabe,Miyahara,Noda}
and asymmetric DOS~\cite{Kanamori,Ulmke,Wahle,BalzerPotthoff}.
On the other hand,
the Stoner theory is inapplicable in the strong coupling regime.
In this limit, itinerant nature of one hole
in the Hubbard model on a lattice with a closed-loop structure
leads to a fully polarized ferromagnetically ordered ground state, which is
so-called Nagaoka ferromagnetism~\cite{Nagaoka}.
However, still controversial how this ordered state relates to the system with the finite hole density.
What is the most important in this strong-coupling region is that
large Coulomb interactions and low energy itinerant properties of electron are necessary to
deal with precisely in an equal footing.
Therefore, it is desired to access the strongly correlated region
by means of reliable techniques.

In our previous paper~\cite{Kamogawa},
we have examined the ferromagnetic instability
in the single-band Hubbard model
by means of dynamical mean-field (DMFT) theory
and the continuous-time quantum Monte Carlo (CTQMC) simulations.
Then, we have clarified that the ferromagnetically ordered state is realized
in the hypercubic lattice, but it does not in the Bethe lattice.
Unlike the weak coupling regime,
our previous results indicates that the noninteracting density of states with
a slower decay in the high-energy region plays an important role
in realizing the ferromagnetically ordered state at low temperatures.
It is also instructive to clarify how magnetic fluctuations are enhanced
in the paramagnetic metallic region.
To this end, we study, in the paper, the magnetic instability
in the single-band Hubbard model on the hypercubic lattice.

We consider the doped single-band Hubbard model, which
should be given as,
\begin{align}
  H =  -t\sum_{\langle i,j,\rangle ,\sigma} (c_{i\sigma}^{\dagger} c_{j\sigma} + \mathrm{h.c.})
        + U \sum_i n_{i\uparrow} n_{i\downarrow} - \sum_{i\sigma} (\mu +\frac{h}{2}\sigma) n_{i\sigma},
\end{align}
where $c_{i\sigma} (c_{i\sigma}^\dag)$ is the annihilation (creation) operator of an electron
with spin $\sigma(=\uparrow, \downarrow)$ at the $i$th site
and $n_{i\sigma}=c_{i\sigma}^\dag c_{i\sigma}$.
$t$ is the hopping integral, $U$ is the on-site Coulomb interaction,
$\mu$ is the chemical potential, and $h$ is the external magnetic field.

To clarify magnetic properties in the Hubbard model on the hypercubic lattice,
we make use of DMFT~\cite{Metzner,Muller,RevModPhys.68.13,Pruschke}.
In DMFT, the lattice model is mapped to a single impurity model connected
dynamically to a ``heat bath''.
The electron Green's function is self-consistently obtained
via this impurity problem.
The treatment is exact in the infinite dimensions and
DMFT has successfully been explained
magnetic properties in single-band~\cite{Chitra_Kotliar_1999,Zitzler2002,Zitzler_Tong_Pruschke_Bulla_2004,Koga_Saitou_Yamamoto_2013,Kamogawa} and multiorbital models~\cite{Momoi_Kubo_1998,Miyatake_Inaba_Suga_2010,Yanatori_Koga_2015,Yanatori_Koga_2016,Koga_Yanatori_2017,Ishigaki_Nasu_Koga_2019,Ishigaki_Nasu_Koga_Hoshino_Werner_2019}.

In DMFT, the selfenergy is represented to be site-diagonal $\Sigma_\sigma(k,z)=\Sigma_\sigma(z)$ and
the lattice Green's function is given as
\begin{eqnarray}
  G_\sigma (k,z)^{-1} &=&   G_{0\sigma} (k,z)^{-1} -\Sigma_\sigma(z),
\end{eqnarray}
where $G_{0\sigma} (k,z)^{-1}=z+ \mu +\frac{h}{2}\sigma-\epsilon_k$,
$\epsilon_k(=-2t\sum_i^d \cos k_i)$ is the dispersion relation, and $d$ is the dimension.
The local Green's function is then obtained as
\begin{eqnarray}
  G_{loc,\sigma} (z) =
  \int dk  G_\sigma (k,z)
  =\int dx \frac{\rho_0(x)}{z+ \mu +\frac{h}{2}\sigma- x - \Sigma_{\sigma} (i \omega_n )},
  \label{rho}
\end{eqnarray}
where we have introduced the non-interacting DOS for the hypercubic lattice
\begin{eqnarray}
  \rho_0(x)=\int dk \delta (x-\epsilon_k) = \frac{1}{\sqrt{\pi}D} \exp\left[ -\left(\frac{x}{D}\right)^2\right],
\end{eqnarray}
where $D(=2\sqrt{d}t)$ is the normalized energy scale
characteristic of the tightbinding model on the hypercubic lattice ($d\rightarrow \infty$).
The Dyson equation in the effective impurity model is given as,
\begin{eqnarray}
  {\cal G}_\sigma (z)^{-1} &=&   G_{imp, \sigma} (z)^{-1} +\Sigma_{imp, \sigma}(z),
\end{eqnarray}
where ${\cal G}(z)$ is the effective bath.
We then obtain the selfenergy and Green function,
solving the effective impurity model.
We iterate the procedure so as to satisfy the selfconsistency conditions
$G_\sigma(z)=G_{imp, \sigma}(z)$ and $\Sigma_\sigma(z)=\Sigma_{imp, \sigma}(z)$
until the desired numerical accuracy is achieved.

In this manuscript, we discuss magnetic properties in the doped single-band Hubbard model,
by calculating the uniform magnetization and magnetic susceptibility.
These are defined as,
\begin{eqnarray}
  m=\frac{1}{2}\Big( \langle n_{\uparrow}\rangle -\langle n_{\downarrow} \rangle\Big),\hspace{5mm}
  \chi=\lim_{\Delta h\rightarrow 0}\frac{\Delta m}{\Delta h},
\end{eqnarray}
where $n_\sigma=\sum_i n_{i\sigma}/N$ and $N$ is the number of sites.
In our study, we numerically evaluate the magnetic susceptibility by
the magnetization induced by a tiny external magnetic field;
$\Delta m$ is given by the difference between $m$ for $h=\Delta h$ and that for $h=0$.
Here,
we study the nature of the ferromagnetic metallic state
in the single-band Hubbard model.
To this end, we do not consider the antiferromagnetically ordered state and phase separation,
which might be realized near half filling~\cite{NCA,Zitzler2002},
where $n=\sum_{\sigma}\langle n_\sigma\rangle$.
This enables us to clarify the essence of
the ferromagnetic instability
in the large-$U$ region.

In our calculations,
we employ the strong-coupling version of the CTQMC method~\cite{PhysRevLett.97.076405,RevModPhys.83.349}
based on the segment algorithm, which is one of the powerful methods
to solve the effective impurity model.
In the method, Monte Carlo samplings are performed
by local updates such as insertion (removal) of a segment or empty space
between segments (antisegment).
However, the acceptance probability for the updates $p_S$
is exponentially suppressed while increasing the interaction strength $U$, typically around $n\sim 1$.
Therefore, it is difficult to evaluate the Green's function
in the reasonable computational cost.
Now, we consider additional updates, where
the configuration for both spins in a certain interval
is simultaneously changed.
This double flip update enables us to perform the CTQMC method
in the strong-coupling region efficiently~\cite{KogaDBLE,Yanatori_Koga_2016,Koga_Yanatori_2017,Kamogawa}.
\begin{figure*}[htb]
\centering
\includegraphics[width=\columnwidth]{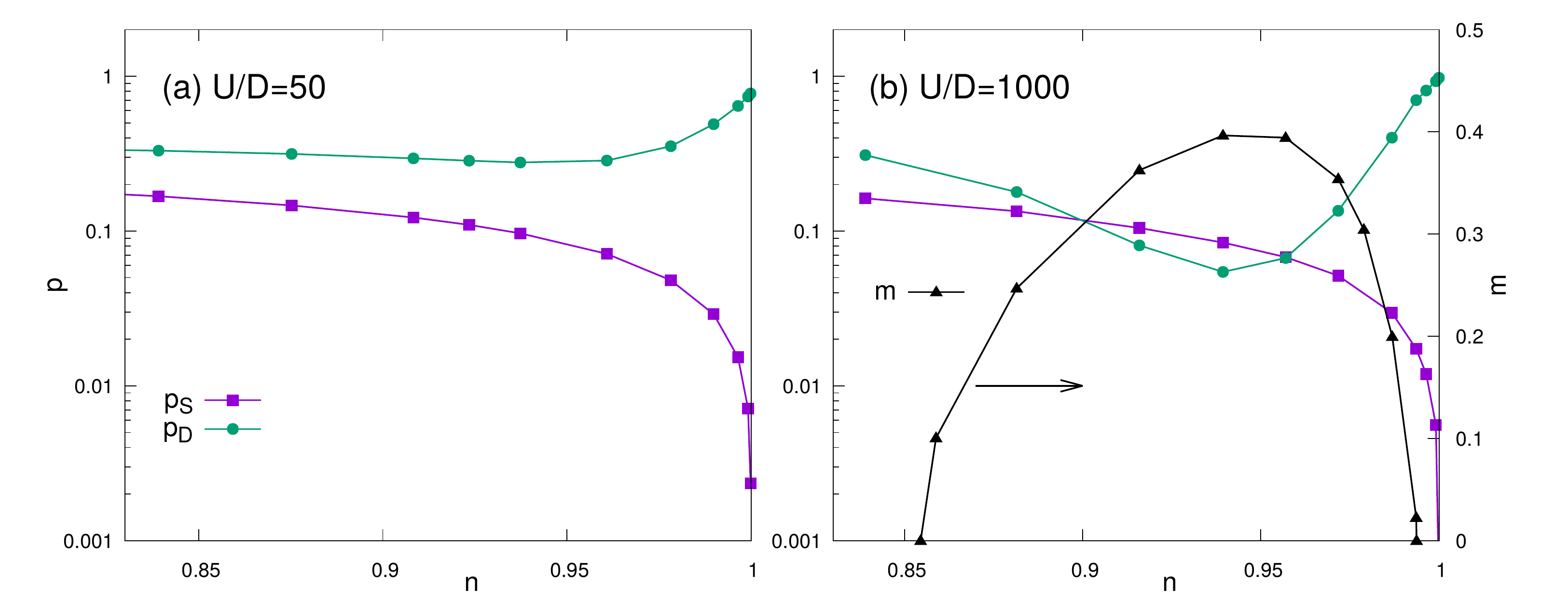}
\caption{
  The acceptance ratios for simple and double flip updates
  as a function of the electron density $n$
  in the system with $U/D=50$ (a) and $U/D=1000$ (b) when $T/D=0.01$.
  Triangles in (b) represent the spontaneous magnetization.
}
\label{fig:acpt}
\end{figure*}
Figure~\ref{fig:acpt} shows the acceptance ratios $p_S$ and $p_D$
for the standard and double flip updates in the case with $U/D=50$ and $1000$
at a fixed temperature $T/D=0.01$.
It is found that around $n\sim 1$,
the double flip update processes are almost accepted $(p_D\sim 1)$, while
the standard updates little occur $(p_S \ll 1)$.
Therefore, we can say that
the double flip updates are necessary to evaluate magnetic quantities accurately.
On the other hand, away from half filling,
the system becomes metallic, where
$p_S$ rapidly increases and $p_D$ decreases.
When $U/D=1000$,
the uniform magnetization appears when $0.85 < n < 0.99$, as shown in Fig.~\ref{fig:acpt}(b).
In the case, the update process for the spin inversion is hard to be accepted, namely $p_D$ is suppressed.

\begin{figure*}[htb]
\centering
\includegraphics[width=\columnwidth]{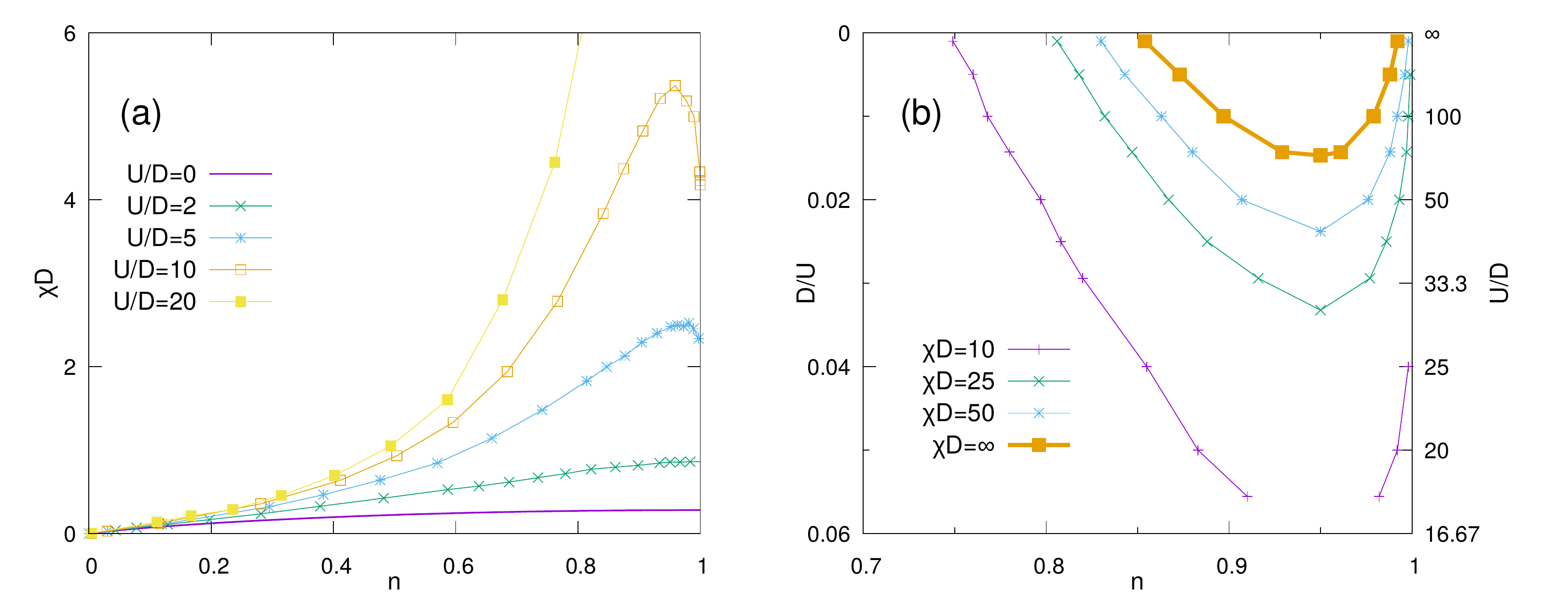}
\caption{
  (a) Magnetic susceptibility as a function of electron filling
  in the system with $U/D=0, 2, 5, 10,$ and $20$ at $T/D=0.01$.
  (b) Contour plot of the magnetic susceptibility
  in the single-band Hubbard model when $T/D=0.01$.
  The bold line represents the phase boundary between the paramagnetic and ferromagnetic phases, which is determined by the divergence of the susceptibility.
}
\label{fig:PD}
\end{figure*}

Using the above update processes in the Monte Carlo simulations,
we study magnetic properties in the single band Hubbard model on the hypercubic lattice.
Figure~\ref{fig:PD}(a) shows the electron density dependence of
the susceptibility
in the weak coupling region at $T/D=0.01$.
When the system is noninteracting $(U=0)$,
the susceptibility is proportional to the noninteracting density of states
at Fermi level.
Therefore, the susceptibility takes a maximum at $n=1$.
The introduction of the Coulomb interaction enhances magnetic fluctuations,
typically, around $n\sim 1$, as seen in Fig.~\ref{fig:PD}(a).
We find that in the case with $U/D=5$, the magnetic susceptibility takes
a maximum away from $n=1$.
This should indicate the ferromagnetic instability away from half filling.
To clarify how the ferromagnetic instability appears
in the strong coupling region,
we show in Fig.~\ref{fig:PD}(b) the contour plot of the magnetic susceptibility
in the model at $T/D=0.01$.
When the interaction strength increases,
nonmonotonic behavior in the susceptibility appears
as a function of the filling $n$.
When the Coulomb interaction is fixed as a certain value $(U/D\lesssim 70)$,
the peak structure in the susceptibility is always located around $n\sim 0.95$,
in contrast to the weak coupling region.
Further increase of the interaction strength $U$ drives the system
to the ferromagnetically ordered state,
where the second-order phase transition occurs with a divergence of
the magnetic susceptibility.
This is in contrast to the single-band Hubbard model on the Bethe lattice.
In the case, the maximum in the susceptibility is always located at $n=1$~\cite{Kamogawa}, which suggests that the ferromagnetic instability is hindered by the antiferromagnetic order.
On the other hand, in the case of the hypercubic lattice, the maximum location away from $n=1$ indicates the appearance of the ferromagnetic order in the system.
It is an interesting problem to clarify
whether or not the ferromagnetic instability is indeed present in the single-band Hubbard model in the finite dimensions, which is now under consideration.


We have studied a ferromagnetic instability in a single-band Hubbard model
on the hypercubic lattice away from half filling,
combining dynamical mean-field theory with
the continuous-time quantum Monte Carlo simulations.
By calculating the magnetic susceptibility systematically,
we have clarified how ferromagnetic fluctuations are enhanced
in the paramagnetic metallic state.

\begin{acknowledgments}
Parts of the numerical calculations were performed
in the supercomputing systems in ISSP, the University of Tokyo.
This work was supported by Grant-in-Aid for Scientific Research from
JSPS, KAKENHI Grant Nos. JP19H05821, JP18K04678, JP17K05536 (A.K.),
JP16H02206, JP18H04223, JP19K03742, and by JST PREST (JPMJPR19L5) (J.N.).
The simulations have been performed using some of the ALPS libraries~\cite{alps2}.
\end{acknowledgments}

\nocite{*}

\bibliography{./refs}

\end{document}